\documentclass[pra,twocolumn,showpacs,superscriptaddress,10pt]{revtex4-1}
\usepackage[colorlinks,citecolor=blue,linkcolor=red,dvipdfm]{hyperref}
\usepackage{amsmath,amssymb,graphicx}
\usepackage{times}

\begin{document}

\title{Quadratic fractional solitons}

\author{Liangwei Zeng}
\affiliation{Shenzhen Key Laboratory of Micro-Nano Photonic Information Technology,
Key Laboratory of Optoelectronic Devices and Systems of Ministry of Education and Guangdong Province,
College of Physics and Optoelectronic Engineering, Shenzhen University, Shenzhen 518060, China}

\author{Yongle Zhu}
\affiliation{Shenzhen Key Laboratory of Micro-Nano Photonic Information Technology,
Key Laboratory of Optoelectronic Devices and Systems of Ministry of Education and Guangdong Province,
College of Physics and Optoelectronic Engineering, Shenzhen University, Shenzhen 518060, China}

\author{Boris A. Malomed}
\affiliation{Department of Physical Electronics, School of Electrical Engineering, Faculty of Engineering,
and Center for Light-Matter Interaction, Tel Aviv University, P.O.B. 39040, Tel Aviv, Israel}
\affiliation{Instituto de Alta Investigaci\'{o}n, Universidad de Tarapac\'{a}, Casilla 7D, Arica, Chile}

\author{Dumitru Mihalache}
\affiliation{Horia Hulubei National Institute of Physics and Nuclear Engineering, Magurele, Bucharest, RO-077125, Romania}

\author{Qing Wang}
\affiliation{College of Science, JiuJiang University, Jiujiang 334000, Jiangxi, China}

\author{Hu Long}
\affiliation{Shenzhen Key Laboratory of Micro-Nano Photonic Information Technology,
Key Laboratory of Optoelectronic Devices and Systems of Ministry of Education and Guangdong Province,
College of Physics and Optoelectronic Engineering, Shenzhen University, Shenzhen 518060, China}

\author{Yi Cai}
\affiliation{Shenzhen Key Laboratory of Micro-Nano Photonic Information Technology,
Key Laboratory of Optoelectronic Devices and Systems of Ministry of Education and Guangdong Province,
College of Physics and Optoelectronic Engineering, Shenzhen University, Shenzhen 518060, China}

\author{Xiaowei Lu}
\affiliation{Shenzhen Key Laboratory of Micro-Nano Photonic Information Technology,
Key Laboratory of Optoelectronic Devices and Systems of Ministry of Education and Guangdong Province,
College of Physics and Optoelectronic Engineering, Shenzhen University, Shenzhen 518060, China}

\author{Jingzhen Li}
\email{lijz@szu.edu.cn}
\affiliation{Shenzhen Key Laboratory of Micro-Nano Photonic Information Technology,
Key Laboratory of Optoelectronic Devices and Systems of Ministry of Education and Guangdong Province,
College of Physics and Optoelectronic Engineering, Shenzhen University, Shenzhen 518060, China}

\begin{abstract}
We introduce a system combining the quadratic self-attractive or composite
quadratic-cubic nonlinearity, acting in the combination with the fractional
diffraction, which is characterized by its L\'{e}vy index $\alpha $. The
model applies to a gas of quantum particles moving by L\'{e}vy flights, with
the quadratic term representing the Lee-Huang-Yang correction to the
mean-field interactions. A family of fundamental solitons is constructed in
a numerical form, while the dependence of its norm on the chemical potential
characteristic is obtained in an exact analytical form. The family of
\textit{quasi-Townes solitons}, appearing in the limit case of $\alpha =1/2$, is
investigated by means of a variational approximation. A nonlinear lattice,
represented by spatially periodical modulation of the quadratic term, is
briefly addressed too. The consideration of the interplay of competing
quadratic (attractive) and cubic (repulsive) terms with a lattice potential
reveals families of single-, double-, and triple-peak gap solitons (GSs) in
two finite bandgaps. The competing nonlinearity gives rise to alternating
regions of stability and instability of the GS, the stability intervals
shrinking with the increase of the number of peaks in the GS.
\end{abstract}

\maketitle

\section{Introduction}

It is commonly known that soliton families are supported by the nonlinear
Schr\"{o}dinger equation (NLSE) \cite{REV1,REV2,REV3,REV4,AVK,REV5,REV6,
REV7,REV8,REV9,REV10,REV11} [alias the
Gross-Pitaevskii equation (GPE) for Bose-Einstein condensates (BECs) \cite%
{Pit,REVBEC}] in a great number of realizations. Still broader varieties of
solitons have been predicted in generalized models of the NLSE\ type, such
as ones with the strength of the repulsive nonlinearity rapidly growing from
the center to periphery \cite{SDN1,SDN2,SDN3,SDN4,SDN5,SDN6,SDN7,SDN8,SDN9}.
Solitons in linear \cite{LL1,LL2,LL3,LL4,LL5,LL6,LL7,Morsch,LL8,LL9,LL10}
and nonlinear spatially periodic potentials \cite{NL1,NL2,NL3}
(alias linear/nonlinear lattices) have also been a subject of
many works. Recently, defects in linear \cite{LL10} and nonlinear \cite{NL4}
lattices have been demonstrated to stabilize different types of solitons.
Linear periodic potentials form spectral bandgaps \cite{GAP1}, which are
populated by gap-soliton (GS)\ families, which include fundamental,
multipole, and vortex modes. GSs have been widely studied theoretically \cite%
{GAP1,LL6,Morsch} and observed in experiments \cite{GAP2,GAP3,Oberthaler}.

The introduction of fractional calculus in NLSE has been drawing much
interest since it was proposed by Laskin (originally, in the linear form) as
the quantum-mechanical model, derived from the respective Feynman-integral
formulation, for particles moving by L\'{e}vy flights \cite%
{Lask1,Lask2,Lask3}. Experimental implementation of fractional linear Schr%
\"{o}dinger equations has been reported in condensed matter \cite{EXP1,EXP2}%
, optics \cite{EXP3} (in the form of transverse dynamics in laser cavities),
and quantum physics \cite{EXP4}. Further, realization of the propagation
dynamics of light beams governed by this equation was proposed \cite{PROP},
and its extension for the model including the $\mathcal{PT}$ symmetry was
put forward too \cite{PTS}. Many types of optical solitons produced by
fractional NLSEs have been theoretically investigated \cite{soliton1}--\cite%
{Frac20}, such as \textquotedblleft accessible solitons" \cite{Frac1,Frac2}%
, GSs \cite{Frac5a,Frac5b,soliton4,Frac5c,Frac5}, solitary vortices \cite{Frac6,Frac7},
multipole and multi-peak solitons \cite{Frac8,Frac9,Frac10,Frac11}, soliton
clusters \cite{Frac12}, symmetry-breaking solitons \cite%
{Frac15,Frac16,Frac17} as well as solitons in couplers \cite{Frac18,Frac19,Frac20}.

Still missing is the analysis of settings combining the fractional
diffraction/dispersion and quadratic\ nonlinearity, except for a brief
mention of the possibility to derive a two-component system of equations for
the second-harmonic generation in the fractional-dimension space \cite%
{Thirouin}. In this work, we introduce a model combining the self-attractive
quadratic nonlinearity, or competing attractive-repulsive quadratic-cubic (QC)
nonlinearity, and the fractional diffraction. A linear lattice potential is
also included, in the general case. The model is an amended GPE for the
description of a quasi-one-dimensional BEC composed of particles moving by L%
\'{e}vy flights, with the cubic nonlinearity representing, as usual, contact
interactions between particles, treated in the mean-field approximation \cite%
{Pit}. The sign of the cubic term may correspond to either repulsive or
attractive inter-particle interactions. The quadratic term represents the
beyond-mean-field correction induced by quantum fluctuations around
mean-field states [the Lee-Huang-Yang (LHY)\ effect \cite{LHY}], in a binary
BEC with contact repulsion between identical particles and attraction
between ones representing different species in the two-component mixture. In
the three-dimensional (3D) form, the LHY correction, in the form of a quartic self-repulsive
term (assuming equal wave functions of the two components), was derived by
Petrov \cite{Petrov} (see also Ref. \cite{HHu}). The accordingly amended GPE
had predicted the existence of stable 3D and quasi-2D self-trapped states in
the form of \textquotedblleft quantum droplets" (QDs), which were quickly
demonstrated experimentally \cite{QD1,QD2,QD3,QD4}. The reduction of the
dimension from $3$ to $1$ leads to the replacement of the repulsive quartic
LHY term by an \emph{attractive} quadratic one \cite{Petr-Astr}, which leads
to the prediction of specific quasi-1D QDs \cite{Astra} and intrinsic modes
in them \cite{Tylutki}. This quadratic term plays the central role in the
analysis presented in this work.

First, we address the basic version of the model, with the quadratic-only
nonlinearity acting in the free space (without the lattice potential),
construct a family of fundamental solitons in it, and identify their
stability region. The consideration of the setting without the cubic term is
relevant because it may be eliminated by means of the Feshbach resonance,
adjusting the strength of the inter-species interaction in the binary BEC
\cite{Feshbach}. The resulting \textit{LHY fluid} has been observed
experimentally \cite{LHY-liquid}. Then, we demonstrate that the interplay of
the general QC nonlinearity, which includes competing self-attractive
quadratic and repulsive cubic terms, with the underlying lattice potential
creates various types of gap solitons, including two- and three-peak ones.
The system with the competing nonlinearities is interesting, as the previous
study of gap solitons in the fractional setting has produced very different
results for the self-repulsive and attractive signs of the cubic term \cite%
{Frac5a}, while competing nonlinearities were not considered there.

The rest of this paper is organized as follows. We formulate the model,
including scaling relations and the variational approximation (VA)
for soliton families in the free-space setting in Sec. \ref{sec2}.
Numerical results for the
basic case of the self-attractive quadratic LHY nonlinearity in the free
space are reported in Sec. \ref{sec3}, which also presents, for the sake of
comparison, similar results based on the cubic self-attraction. In addition,
the same Section briefly addresses solitons supported by the quadratic
nonlinear lattice, with the self-attraction coefficient subject to spatially
periodic modulation. Results for families of single-, double-, and
triple-peak GSs, supported by the competing QC nonlinearity, combined with
the lattice potential, are summarized in Sec. \ref{sec4}. Finally,
the paper is concluded by Sec. \ref{sec5}.

\section{The model and variational approximation}
\label{sec2}

\subsection{Basic equations}

In the scaled form, the LHY-amended fractional GPE, for identical wave
functions $\Psi $ of the two components of the binary BEC composed of particles
moving by L\'{e}vy flights, is written as
\begin{equation}
i\frac{\partial \Psi }{\partial t}=\frac{1}{2}\left( -\frac{\partial ^{2}}{%
\partial x^{2}}\right) ^{\alpha /2}\Psi +V(x)\Psi +\xi \left\vert \Psi
\right\vert \Psi +\mathrm{g}\left\vert \Psi \right\vert ^{2}\Psi .
\label{NLFSE}
\end{equation}%
Here $t$ and $x$ are the normalized time and coordinate, while $\xi <0$ and $%
\mathrm{g}>0/\mathrm{g}<0$ are the coefficients of the attractive quadratic
(LHY) and repulsive/attractive cubic (mean-field) nonlinearities,
respectively. The fractional-diffraction operator with L\'{e}vy index (LI) $%
\alpha $ is defined, as usual, by means of the Fourier transform (also called
the Riesz derivative):
\cite{Lask1,Lask2,Lask3,EXP1,EXP2,EXP3,EXP4,soliton2}:
\begin{equation}
\left( -\frac{\partial ^{2}}{\partial x^{2}}\right) ^{\alpha /2}\Psi =\frac{1%
}{2\pi }\int_{-\infty }^{+\infty }ds|s|^{\alpha }\int_{-\infty }^{+\infty
}d\eta e^{is(x-\eta )}\Psi (\eta ),  \label{FracDefi}
\end{equation}
the normal diffraction corresponding to $\alpha =2$. The lattice potential
in Eq. (\ref{NLFSE}) with strength $V_{0}$ is taken as
\begin{equation}
V=V_{0}~\mathrm{sin}^{2}x  \label{LPE}
\end{equation}%
(in the free space, $V_{0}=0$).

We look for stationary solutions to Eq. (\ref{NLFSE}) with a real chemical
potential $\mu $ in the form of
\begin{equation}
\Psi =U(x)\mathrm{exp}(-i\mu z).  \label{SSE}
\end{equation}%
The substitution of $\Psi$ in Eq. (\ref{NLFSE}), leads to the stationary
equation,
\begin{equation}
\mu U=\frac{1}{2}\left( -\frac{\partial ^{2}}{\partial x^{2}}\right)
^{\alpha /2}U+V(x)U+\xi \left\vert U\right\vert U+\mathrm{g}\left\vert
U\right\vert ^{2}U.  \label{NLFSES}
\end{equation}%
The stationary localized states are characterized by their norm,
\begin{equation}
N=\int_{-\infty }^{+\infty }\left\vert U(x)\right\vert ^{2}dx.  \label{P}
\end{equation}

To introduce the linear-stability analysis for the stationary states, the
perturbed solution is defined as
\begin{equation}
\Psi =[U(x)+p(x)\mathrm{exp}(\lambda t)+q^{\ast }(x)\mathrm{exp}(\lambda
^{\ast }t)]\mathrm{exp}(-i\mu t),  \label{PERB}
\end{equation}%
where $p(x)$ and $q^{\ast }(x)$ are small complex perturbations (the
asterisk denotes complex conjugation), and $\lambda $ is the instability growth
rate. Substituting this ansatz in Eq. (\ref{NLFSE}), one derives the
linearized Bogoliubov-de Gennes equations for $p$ and $q$:
\begin{equation}
\left\{ \begin{aligned}
i\lambda p=&+\frac{1}{2}\left(-\frac{\partial^2}{\partial
x^2}\right)^{\alpha/2}p+(V-\mu)p+\mathrm{g}U^2(2p+q)\\
&+\frac{1}{2}\xi|U|(3p+q), \\
i\lambda q=&-\frac{1}{2}\left(-\frac{\partial^2}{\partial
x^2}\right)^{\alpha/2}q+(\mu-V)q-\mathrm{g}U^2(2q+p)\\
&-\frac{1}{2}\xi|U|(3q+p). \end{aligned}\right.  \label{LAS}
\end{equation}%
This form of the equations is valid only in the case when the underlying
stationary solution $U(x)$ is a real one [the presence of $|U|$ in Eq. (\ref%
{LAS}) is relevant if $U(x)$ is a sign-changing function]. The underlying
stationary solution (\ref{SSE}) is stable solely in the case when Eq. (\ref%
{LAS}) produces purely imaginary eigenvalues $\lambda $.

\subsection{Scaling relations for soliton families}

In the case of the attractive cubic-only nonlinearity in the free space,
with
\begin{equation}
\xi =V=0,\mathrm{g}<0,  \label{cubic only}
\end{equation}%
Eq. (\ref{NLFSES}) gives rise to the exact scaling relation between the
soliton's norm and chemical potential:%
\begin{equation}
N_{\mathrm{cubic}}(\mu ;\mathrm{g})=N_{\mathrm{cubic}}^{(0)}(\alpha )|%
\mathrm{g}|^{-1}\left( -\mu \right) ^{1-1/\alpha },  \label{cubic}
\end{equation}%
with a constant $N_{\mathrm{cubic}}^{(0)}(\alpha )$ [for the usual NLSE, it
is $N_{\mathrm{cubic}}^{(0)}(\alpha =2)=2\sqrt{2}$, see Eq. (\ref%
{cubic-usual}) below]. The fact that this relation satisfies the commonly
known Vakhitov-Kolokolov (VK) criterion, $dN/d\mu <0$ \cite{VK,Berge}, at $%
\alpha >1$, suggests that the respective soliton family may be stable, see
further details in Figs. \ref{fig3} and \ref{fig4} below. The case of $%
\alpha =1$, which corresponds to the degenerate form of relation (\ref{cubic}%
), with $N_{\mathrm{cubic}}(\mu )\equiv \mathrm{const}$, implies the
occurrence of the \textit{critical collapse}, which makes all solitons
unstable (cf. the commonly known cubic NLSE in the 2D space with the normal
diffraction, $\alpha =2$, in which the family of \textit{Townes' solitons},
destabilized by the critical collapse, have a constant norm \cite{Berge}).
In the case of $\alpha <1$, the solitons generated by Eq. (\ref{NLFSES})
with $\xi =V=0$ and $\mathrm{g}<0$ are strongly destabilized by the presence
of the \textit{supercritical collapse}, similar to the solitons of the usual
cubic NLSE in three dimensions \cite{Berge}.

In the case of the quadratic self-attraction in the free space,
\begin{equation}
\mathrm{g}=V=0,\xi <0,  \label{quadr only}
\end{equation}%
a scaling relation for the soliton's norm, similar to Eq. (\ref{cubic}), is
\begin{equation}
N_{\mathrm{quadr}}(\mu ;\xi )=N_{\mathrm{quadr}}^{(0)}(\alpha )|\xi
|^{-2}\left( -\mu \right) ^{2-1/\alpha },  \label{quadr}
\end{equation}%
with a constant $N_{\mathrm{quadr}}^{(0)}(\mu )$ [in the case of the usual
diffraction, $N_{\mathrm{quadr}}^{(0)}(\alpha =2)=3\sqrt{2}$, see Eq. (\ref%
{quadr-usual}) below]. Equation (\ref{quadr}) demonstrates that the
respective solitons satisfy the VK stability criterion at $\alpha >1/2$.
Equation (\ref{NLFSES}) with the quadratic nonlinearity gives rise to the
degenerate family of \textit{quasi-Townes} solitons with a constant norm,
which are destabilized by the presence of the critical collapse, at $\alpha
=1/2$. The subcritical collapse takes place at $\alpha <1/2$.

In the QC model including both $\mathrm{g}<0$ and $\xi <0$ in Eq. (\ref%
{NLFSES}), the scaling relations (\ref{cubic}) and (\ref{quadr}) dominate,
respectively, at $|\mu |\rightarrow \infty $ and $|\mu |\rightarrow 0$. In
the latter limit, the relation (\ref{quadr}) is relevant too for the QC model
with the competing quadratic self-attraction and cubic repulsion, i.e., $\xi
<0$ and $\mathrm{g}>0$. In the system with the competing nonlinearities, the
solitons exist in the interval of the chemical potential that was
identified for QDs in Ref. \cite{Astra}:%
\begin{equation}
\mu _{\min }\equiv -2\xi ^{2}/\left( 9\mathrm{g}\right) <\mu <0.  \label{min}
\end{equation}%
In the limit of $\mu \rightarrow \mu _{\min }$, the solitons develop the
shape of flat-top QDs (see Ref. \cite{Astra}), with the local density
approaching the limit value,
\begin{equation}
U_{\max }^{2}=\left( 2\xi /3\mathrm{g}\right) ^{2},
\end{equation}%
and a logarithmically large length,%
\begin{equation}
l\simeq \left( -2\mu \right) ^{1/\alpha }\ln \left( \left( \mu -\mu _{\min
}\right) ^{-1}\right) .  \label{L}
\end{equation}%
Such flat-top solitons are stable, and they support a large number of
oscillatory internal modes \cite{Tylutki}.

\begin{figure*}[tbp]
\begin{center}
\includegraphics[width=1.6\columnwidth]{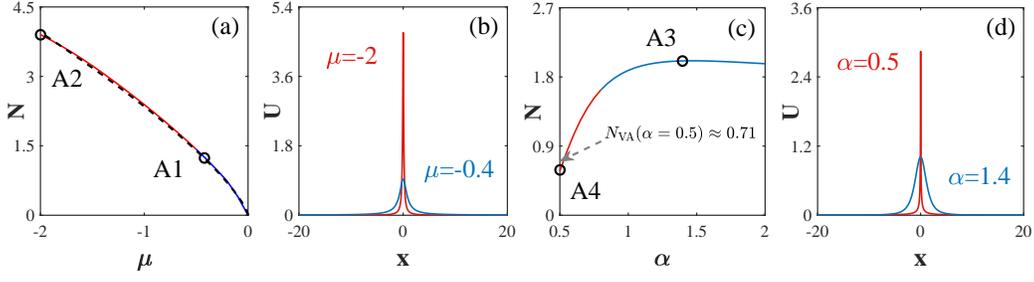}
\end{center}
\caption{(a) The dependence $N(\protect\mu )$ and (b) the profiles of solitons
labeled A1 and A2 in (a) (corresponding to $\protect\mu =-0.4$ and $-2$,
respectively) for the soliton family produced by the numerical solution of
Eq. (\protect\ref{NLFSES}) with the quadratic-only nonlinearity ($\protect%
\xi =-1,\mathrm{g}=0$) in the free space ($V=0$), at a fixed LI, $\protect%
\alpha =0.8$. (c) The dependence $N(\protect\alpha )$ and (d) the profiles of
solitons labeled A3 and A4 (corresponding to $\protect\alpha =1.4$ and $0.5$%
, respectively) for a fixed chemical potential, $\protect\mu =-0.6$. The blue
and red segments of the curves in panels (a) and (c) denote subfamilies of
stable and (weakly) unstable solitons, respectively. The black dashed line in (a)
represents the analytical scaling relation (\protect\ref{quadr}). The value
marked by the gray dashed arrow is the one predicted by Eq. (\protect\ref%
{quadr-VA}), its numerically found counterpart being $N_{\mathrm{numer}}(%
\protect\alpha =0.5)\approx 0.57$. The evolution of the solitons labeled by
A1--A4 are displayed in Fig. \protect\ref{fig2}.}
\label{fig1}
\end{figure*}

\begin{figure*}[tbp]
\begin{center}
\includegraphics[width=1.6\columnwidth]{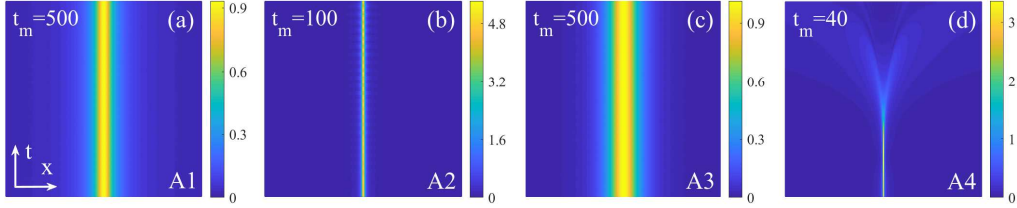}
\end{center}
\caption{The perturbed evolution of the solitons corresponding to labels
A1--A4 in Figs. \protect\ref{fig1}(a,c). The respective values of the
parameters are $\protect\alpha =0.8$, $\protect\mu =-0.4$ (a); $\protect%
\alpha =0.8$, $\protect\mu =-2$ (b); $\protect\alpha =1.4$, $\protect\mu %
=-0.6$ (c); $\protect\alpha =0.5$, $\protect\mu =-0.6$ (d). Other parameters
are $\protect\xi =-1$, $\mathrm{g}=V_{0}=0$. In all panels, the evolution is
plotted in the interval $-10<x<+10$, while the values $\mathrm{t}_{\mathrm{m}}$
indicate time intervals in the respective panels, $0<t<\mathrm{t}_{\mathrm{m}%
}$. The value of $\mathrm{t}_{\mathrm{m}}$\ is taken large in panels (a) and
(c), to verify the complete stability of the respective solitons.}
\label{fig2}
\end{figure*}

\subsection{The variational approximation (VA)}

In the context of a different problem, it was recently demonstrated that VA
can be used to look for localized solutions of the fractal NLSE \cite%
{Frac14,we}. To this end, we note that Eq. (\ref{NLFSES}) for real $U(x)$,
with the fractional diffraction operator defined as per Eq. (\ref{FracDefi})
can be derived from the Lagrangian (see also Ref. \cite{Baleanu}),
\begin{equation}
\begin{aligned}
L=&\frac{1}{8\pi }\int_{-\infty }^{+\infty }ds|s|^{\alpha }\int_{-\infty
}^{+\infty }d\eta \int_{-\infty }^{+\infty }dxe^{is(x-\eta )}U(x)U(\eta ) \\
&+\frac{1}{2}\int_{-\infty }^{+\infty }dx\left[ -\mu +V(x)\right] U^{2}(x)\\
&+\frac{\xi }{3}\int_{-\infty }^{+\infty }dxU^{3}(x)+\frac{\mathrm{g}}{4}%
\int_{-\infty }^{+\infty }dxU^{4}(x).  \label{Lagr}
\end{aligned}
\end{equation}
A natural form of the ansatz approximating the solitons sought for is based
on the Gaussian,%
\begin{equation}
\mathcal{U}(x)=A\exp \left( -\frac{x^{2}}{2W^{2}}\right) ,  \label{ans}
\end{equation}%
with real amplitude $A$, width $W$, and the norm calculated as per Eq. (\ref%
{P}),%
\begin{equation}
\mathcal{N}=\sqrt{\pi }A^{2}W  \label{Nans}
\end{equation}%
(the calligraphic font is used to denote quantities pertaining to the VA).
The substitution of the ansatz in Lagrangian (\ref{Lagr}) yields the
corresponding effective Lagrangian, which is written here with the amplitude $A$
replaced by the norm, pursuant to Eq. (\ref{Nans}), and the lattice
potential taken as per Eq. (\ref{LPE}):
\begin{eqnarray}
L_{\mathrm{eff}} &=&\frac{1}{2}\left( \frac{V_{0}}{2}-\mu \right) \mathcal{N}%
-\frac{V_{0}}{4}\mathcal{N}e^{-W^{2}}+\frac{\Gamma \left( \left( 1+\alpha
\right) /2\right) }{4\sqrt{\pi }}\frac{\mathcal{N}}{W^{\alpha }}  \notag \\
&&+\frac{\sqrt{2}\xi }{3\sqrt{3}\pi ^{1/4}}\frac{\mathcal{N}^{3/2}}{\sqrt{W}}%
+\frac{\mathrm{g}}{4\sqrt{2\pi }}\frac{\mathcal{N}^{2}}{W},  \label{Leff}
\end{eqnarray}%
where $\Gamma $ is the Gamma-function. Then, values of $\mathcal{N}$ and $W$
are predicted by the Euler-Lagrange equations,%
\begin{equation}
\partial L_{\mathrm{eff}}/\partial \mathcal{N}=\partial L_{\mathrm{eff}%
}/\partial W=0.  \label{EL}
\end{equation}

The most essential result of the VA is the prediction of the fixed value of
the norm for the \textit{quasi-Townes} solitons, for which the critical
collapse takes place in the free space ($V_{0}=0$) at $\alpha =1/2$ and $%
\alpha =1$ in the cases defined, respectively, by Eqs. (\ref{quadr only})
and (\ref{cubic only}), i.e., with the quadratic-only or cubic-only
nonlinearity:%
\begin{gather}
\left( \mathcal{N}_{\mathrm{Townes}}^{\mathrm{(quadr)}}\right) _{\mathrm{VA}%
}=\frac{27}{32\sqrt{\pi }}\Gamma ^{2}\left( \frac{3}{4}\right) \approx 0.71,
\label{quadr-VA} \\
\left( \mathcal{N}_{\mathrm{Townes}}^{\mathrm{(cubic)}}\right) _{\mathrm{VA}%
}=\sqrt{2}\approx 1.41  \label{cubic-VA}
\end{gather}%
[Eq. (\ref{cubic-VA}) is tantamount to a result recently reported for the
cubic nonlinearity with $\alpha =1$ in Ref. \cite{Frac14}]. These results
may be likened to the well-known VA prediction for the norm of the Townes'
solitons in the 2D NLSE with the cubic self-focusing \cite{Anderson}: $%
\left( \mathcal{N}_{\mathrm{Townes}}^{\mathrm{(2D-cubic)}}\right) _{\mathrm{%
VA}}=2\pi $, while the respective numerical value is $N_{%
\mathrm{Townes}}^{\mathrm{(2D-cubic)}}\approx 5.85$, the relative error of
the VA being $\approx 7\%$. It is shown below in Figs. \ref{fig1}(c) and %
\ref{fig3}(c) that numerically found counterparts of values (\ref{quadr-VA}%
) and (\ref{cubic-VA}) are%
\begin{gather}
\left( N_{\mathrm{Townes}}^{\mathrm{(quadr)}}\right) _{\mathrm{num}}\approx
0.57,  \label{quadr-num} \\
\left( N_{\mathrm{Townes}}^{\mathrm{(cubic)}}\right) _{\mathrm{num}}\approx
1.23.  \label{cubic-num}
\end{gather}%
As well as in the above-mentioned case, the VA predicts the Townes' norms
that are somewhat larger than their numerically found counterparts, with
the error $\approx 20\%$ and $13\%$ for the quadratic and cubic
nonlinearities, respectively. A relatively large size of the error is a
consequence of the complex structure of the nonlinear equation with the
fractional diffraction.

The VA also predicts if the dependence $N(\alpha )$ for a fixed value of $\mu $
is growing or decaying. To this end, we note that, at $\alpha =2$, the usual
NLSE with the quadratic or cubic nonlinearity gives rise to the following $%
N(\mu )$ dependences for the commonly known soliton solutions:%
\begin{gather}
N_{\alpha =2}^{\mathrm{(quadr)}}=3\sqrt{2}\left( -\mu \right) ^{3/2},
\label{quadr-usual} \\
N_{\alpha =2}^{\mathrm{(cubic)}}=2\sqrt{-2\mu }.  \label{cubic-usual}
\end{gather}%
Comparing these with values (\ref{quadr-VA}) and (\ref{cubic-VA}), one
arrives at the conclusion that, in the case of the quadratic-only
nonlinearity, the $N(\alpha )$ dependence is growing at $-\mu >-\mu _{%
\mathrm{crit}}^{\mathrm{(quadr)}}\approx 0.305$, and is decaying at $-\mu <-\mu
_{\mathrm{crit}}$. Similarly, $-\mu _{\mathrm{crit}}^{\mathrm{(cubic)}}=1/4$
for the cubic-only term. These predictions agree with the numerical results
displayed below in Figs. \ref{fig1}(c) and \ref{fig3}(c), respectively.

\begin{figure*}[tbp]
\begin{center}
\includegraphics[width=1.6\columnwidth]{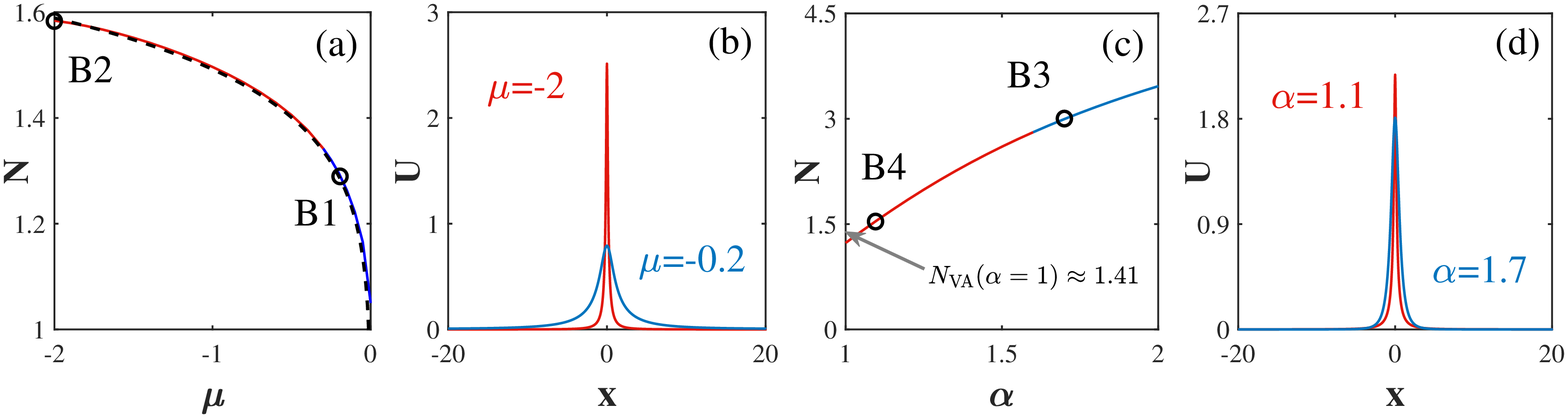}
\end{center}
\caption{(a) The dependence $N(\protect\mu )$ and (b) the profiles of solitons
labeled B1 and B2 in (a) (corresponding to $\protect\mu =-0.2$ and $-2$,
respectively) for the soliton family produced by the numerical solution of
Eq. (\protect\ref{NLFSES}) with the cubic-only nonlinearity ($\protect\xi =0,%
\mathrm{g}=-1$) in the free space ($V=0$), at a fixed LI, $\protect\alpha %
=1.1$. (c) The dependence $N(\protect\alpha )$ and (d) the profiles of solitons
labeled B3 and B4 (corresponding to $\protect\alpha =1.7$ and $1.1$,
respectively) for a fixed chemical potential, $\protect\mu =-1.5$. The blue and
red segments of the curves in panels (a) and (c) denote subfamilies of
stable and (weakly) unstable solitons, respectively. The black dashed line in (a)
represents the analytical scaling relation (\protect\ref{cubic}). The value
marked by the gray dashed arrow is predicted by\ the VA, as per Eq. (\protect
\ref{cubic-VA}). Its numerically found counterpart is $N_{\mathrm{numer}}(%
\protect\alpha =1)\approx 1.41$. The evolution of the solitons labeled by
B1--B4 is displayed in Fig. \protect\ref{fig4}.}
\label{fig3}
\end{figure*}

\begin{figure*}[tbp]
\begin{center}
\includegraphics[width=1.6\columnwidth]{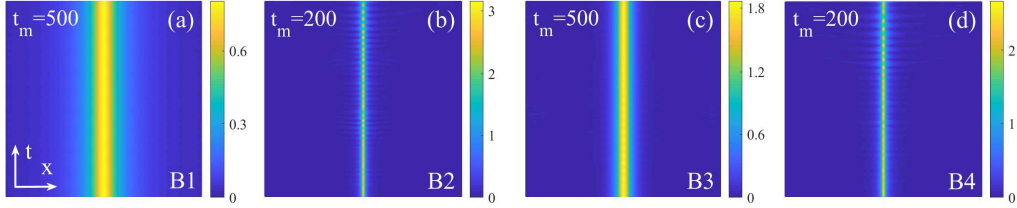}
\end{center}
\caption{The perturbed evolution of the solitons corresponding to labels
B1--B4 in Figs. \protect\ref{fig3}(a,c). The respective values of the
parameters are $\protect\alpha =1.1$, $\protect\mu =-0.2$ (a); $\protect%
\alpha =1.1$, $\protect\mu =-2$ (b); $\protect\alpha =1.7$, $\protect\mu %
=-1.5$ (c); and $\protect\alpha =1.1$, $\protect\mu =-1.5$ (d). Other
parameters are $\mathrm{g}=-1$, $\protect\xi =V_{0}=0$. In all panels, the
evolution is plotted in interval $-10<x<+10$. The values $\mathrm{t}_{\mathrm{m}}
$ indicate time intervals in the respective panels, $0<t<\mathrm{t}_{\mathrm{%
m}}$.}
\label{fig4}
\end{figure*}

Finally, it is relevant to mention that the inclusion of the external
potential, such as the periodic one given by Eq. (\ref{LPE}), helps to
suppress the collapse in the fractional NLSE, and thus to extend the range
of the existence of solitons \cite{Frac9}. In the present case, this
possibility can be illustrated by results following from Eqs. (\ref{Leff})
and (\ref{EL}), which include the weak potential with small $V_{0}>0$, for
the model with the quadratic nonlinearity ($\xi =-1$, $\mathrm{g}=0$).
Namely, at $\alpha =1/2$ the degeneracy of the soliton's norm is (slightly)
lifted by the lattice potential, allowing the solitons to attain the
smallest norm
\begin{equation}
\begin{aligned}
&\left( \mathcal{N}_{\mathrm{\min }}^{\mathrm{(quadr)}}(\alpha =1/2)\right) _{\mathrm{VA}} \\
&\approx \left( \mathcal{N}_{\mathrm{Townes}}^{\mathrm{(quadr)}}\right) _{\mathrm{VA}}
-\frac{27(5/e)^{5/4}}{16\sqrt{2}}\Gamma \left( \frac{3}{4}\right) V_{0} \\
&\approx 0.71-3.13V_{0},  \label{correction to N}
\end{aligned}
\end{equation}
cf. Eq. (\ref{quadr-VA}). Further, the value of $\alpha $ at which the
critical collapse takes place (i.e., the value above which solitons exist)
is lowered by the weak lattice potential from $1/2$ to%
\begin{equation}
\begin{aligned}
\alpha _{\mathrm{crit}}^{\mathrm{(quadr)}}
&\approx \frac{1}{2}-\frac{2\sqrt{2\pi }}{\Gamma (3/4)}(3-\sqrt{5})\sqrt{\sqrt{5}-2}e^{-\left( \sqrt{5}-2\right) /2}V_{0} \\
&\approx \frac{1}{2}-1.35V_{0}.  \label{correction to alpha}
\end{aligned}
\end{equation}
A detailed analysis of these points will be presented elsewhere.

\section{Quadratic solitons in the free space}
\label{sec3}

The stationary solutions of Eq. (\ref{NLFSES}) were constructed by means of the
modified squared-operator method \cite{MSOM}. Then, the stability domains for
such stationary states were identified by eigenvalues $\lambda $, which were
obtained from Eq. (\ref{LAS}) with the help of the Fourier collocation
method \cite{MSOM}. Finally, the predictions for the stability were verified
by direct numerical simulations of the evolution of perturbed solutions in the
framework of Eqs. (\ref{NLFSE}) and (\ref{FracDefi}), using the split-step
fast-Fourier-transform method.
The spatial integration domain we use in this paper is $x\in[-20,20]$, with a spatial step $dx=0.05$ and a time step $dt=0.0005$. The Laplacian-zero boundary condition $(-\partial^{2}/\partial x^{2})^{\alpha /2}\Psi=0$ (at $x=x_{\rm min,max}$) \cite{BC1,BC2} was employed for this work ($x_{\rm min,max}$ represent the left and right end point of integration domain, respectively), and we also obtained the same results by using the Neumann boundary condition. Rerunning the simulations in a larger domain, and/or with smaller $dx$ and $dt$ did not produce any conspicuous change in the pictures.

\subsection{Soliton families supported by the spatially uniform quadratic
and cubic nonlinearities}

\begin{figure*}[tbp]
\begin{center}
\includegraphics[width=2\columnwidth]{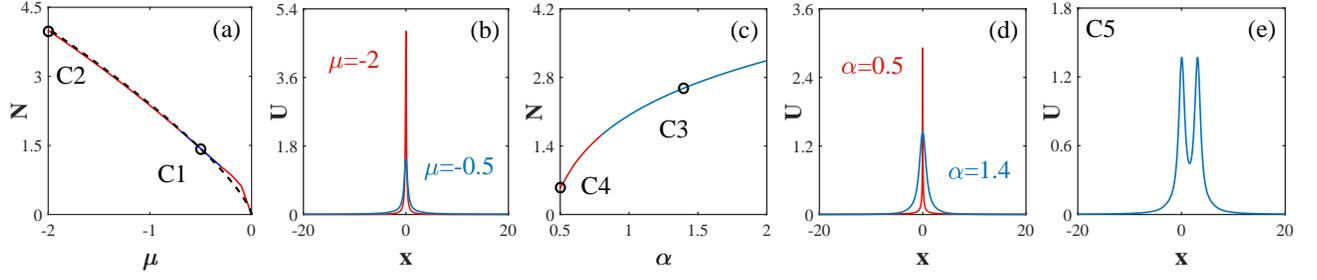}
\end{center}
\caption{(a) The dependence $N(\protect\mu )$ and (b) the profiles of solitons
labeled C1 and C2 in (a) (corresponding to $\protect\mu =-0.5$ and $-2$,
respectively) for the soliton family produced by the numerical solution of
Eq. (\protect\ref{NLFSES}) with the quadratic nonlinearity subject to the
spatially periodic modulations as per Eq. (\protect\ref{quadr}), in the
absence of the linear potential ($V=0$), for a fixed LI, $\protect\alpha %
=0.8 $. (c) The dependence $N(\protect\alpha )$ and (d) the profiles of solitons
labeled C3 and C4 (corresponding to $\protect\alpha =1.4$ and $0.5$,
respectively) for a fixed chemical potential, $\protect\mu =-0.6$. The blue and
red segments in panels (a) and (c) denote subfamilies of stable and (weakly) unstable
solitons, respectively. The black dashed line in (a) represents the
analytical scaling relation (\protect\ref{cubic}), which corresponds to $%
\protect\xi (x)$ replaced by its mean value, $\left\langle \protect\xi %
(x)\right\rangle =-0.5$ [see Eq. (\protect\ref{quadr})]. The evolution of
the solitons labeled by C1--C4 is displayed in Fig. \protect\ref{fig6}. (e)
The shape of the two-peak soliton corresponding to $\protect\alpha =1.2$
and $\protect\mu =-0.6$. Its perturbed evolution is displayed in Fig.
\protect\ref{fig6}(e).}
\label{fig5}
\end{figure*}

\begin{figure*}[tbp]
\begin{center}
\includegraphics[width=2\columnwidth]{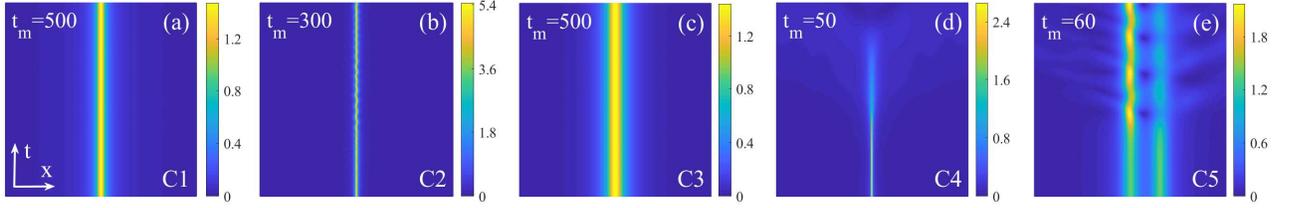}
\end{center}
\caption{The perturbed evolution of the solitons corresponding to labels
C1--C4 in Figs. \protect\ref{fig5}(a,c). The respective values of the
parameters are $\protect\alpha =0.8$, $\protect\mu =-2$ (a); $\protect\alpha %
=0.8$, $\protect\mu =-2$ (b); $\protect\alpha =1.4$, $\protect\mu =-0.6$
(c); $\protect\alpha =0.5$, $\protect\mu =-0.6$ (d). Panel (e) displays the
perturbed evolution of the double-peak soliton whose stationary shape is
presented in Fig. \protect\ref{fig5}(e), for $\protect\alpha =1.2$, $%
\protect\mu =-0.6$. Other parameters are given by Eq. (\protect\ref{quadr}).
In all panels, the evolution is plotted in interval $-10<x<+10$, while the
values $\mathrm{t}_{\mathrm{m}}$ indicate time intervals in the respective
panels, $0<t<\mathrm{t}_{\mathrm{m}}$.}
\label{fig6}
\end{figure*}

Results for solitons produced by Eqs. (\ref{NLFSE}) and (\ref{NLFSES}) in
the basic case of the coefficients taken as per Eq. (\ref{quadr only}) are
displayed in Figs. \ref{fig1} and \ref{fig2}. The soliton families are
characterized by the dependence of norm $N$ on chemical potential $\mu $ for
fixed LI $\alpha $, which is presented in Fig. \ref{fig1}(a) for $\alpha
=0.8$. It is seen that the numerically computed dependence exactly follows
the analytical form given by Eq. (\ref{quadr}), with properly adjusted
constant $N_{\mathrm{quadr}}^{(0)}(\alpha )$. Blue and red segments in the
$N(\mu )$ curve identify, respectively, stable and unstable soliton
subfamilies. Similar results are produced by the numerical solution for
other values of LI from the relevant interval, $0.5<\alpha \leq 2$.

Further, a typical dependence $N(\alpha )$ for a fixed chemical potential, $%
\mu =-0.5$, which is also split in stable and unstable segments, is
presented in Fig. \ref{fig1}(c). Unlike the $N(\mu )$ dependence, this one
cannot be predicted in an exact analytical form. As mentioned above, the VA
based on ansatz (\ref{ans}) predicts the approximate value given by Eq. (\ref%
{quadr-VA}) for the degenerate norm of the quasi-Townes solitons, which is
relatively close to its numerical counterpart (\ref{quadr-num}). At $\alpha
=2$, the numerical value $N_{\mathrm{numer}}\left( \alpha =2\right) \approx
1.97$ is in complete agreement with the exact analytical value given by Eq. (%
\ref{quadr-usual}).

Typical profiles of the solitons with different values of $\mu $ and $\alpha
$, labeled A1--A4 in Figs. \ref{fig1}(a,c), are presented in Figs. %
\ref{fig1}(b,d) respectively. The fact that the soliton becomes narrower
with the increase of $|\mu |$ is another manifestation of the scaling
expressed by Eq. (\ref{quadr}).

\begin{figure*}[tbp]
\begin{center}
\includegraphics[width=1.6\columnwidth]{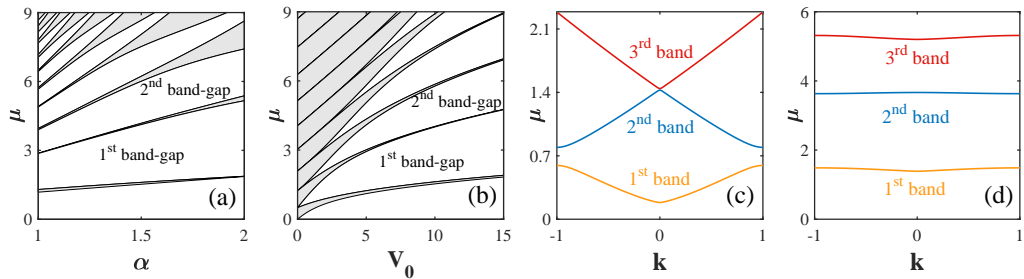}
\end{center}
\caption{The bandgap spectrum produced by the linear version of Eq. (\protect
\ref{NLFSE}): (a) for different values of the LI $\protect\alpha $ at $V_{0}=8$;
(b) for different values of $V_{0}$ at $\protect\alpha =1.3$. The Bloch-wave
spectra obtained for different values of $V_{0}$ at $\protect\alpha =1.3$:
(c) $V_{0}=0.4$ (a weak lattice potential); (d) $V_{0}=8$ (a strong
lattice). }
\label{fig7}
\end{figure*}

The instability of the solitons belonging to the red segments of the $N(\mu
) $ and $N(\alpha )$ dependences in Figs. \ref{fig1}(a,c) is accounted for
by a pair of real eigenvalues $\pm \lambda $. For instance, this pair is $%
\lambda \approx \pm 0.5$ for the weakly unstable soliton shown in Fig. \ref%
{fig2}(b). The perturbed stable and unstable propagation of the solitons
marked A1--A4 in Figs. \ref{fig1}(a,c), whose stationary shapes are shown
in Figs. \ref{fig1}(b,d), is displayed in Fig. \ref{fig2}. It is seen that
the typical instability is quite weak, giving rise [e.g., in Fig. \ref{fig2}%
(b)] to small-amplitude intrinsic vibrations of the soliton and emission of
virtually invisible linear waves. The instability is very different for the
soliton of the Townes type, labeled A4, as seen in Fig. \ref{fig2}(d). It
leads to quick decay of the soliton, which is typical for solitons of this
type \cite{Berge}.

For comparison with the results summarized in Figs. \ref{fig1} and \ref%
{fig2}, systematic findings for solitons supported by the cubic-only
nonlinearity in the free space ($V=0$) are displayed in Figs. \ref{fig3}
and \ref{fig4} (similar results, but produced by the NLSE
with the fractional equation of the Caputo type instead of the Riesz one, were
reported in Ref. \cite{Chen}). In particular, the dependence $N(\mu )$ for such solitons,
displayed in Fig. \ref{fig3}(a), completely agrees with the analytically
predicted scaling given by Eq. (\ref{cubic}). The relationship between $N$
and $\alpha $ for such solitons is presented in Fig. \ref{fig3}(c). As
shown in the latter panel, the VA prediction for the norm of the
quasi-Townes solitons in the case of the cubic nonlinearity at $\alpha =1$,
given by Eq. (\ref{cubic-VA}), is quite close to its numerical counterpart (%
\ref{cubic-num}). At $\alpha =2$, the numerically found norm, $N_{\mathrm{%
numer}}\left( \alpha =2\right) \approx 3.46$, is in complete agreement with
the exact analytical value given by Eq. (\ref{cubic-usual}).

Profiles of solitons labeled B1--B4 in Figs. \ref{fig3}(a,c) are presented
in Figs. \ref{fig3}(b,d), respectively. The perturbed evolution of the
stable and unstable solitons is reported, severally, in Figs. \ref{fig4}%
(a,c) and \ref{fig4}(b,d). It is seen that, in this case too, the instability
is relatively weak. It excites intrinsic vibrations of the solitons, without
destroying them.

In the case of the mixed QC nonlinearity in the free space, the results are
qualitatively similar to those displayed in Figs. \ref{fig3} and \ref{fig4}%
. In particular, the mixture of the quadratic and cubic self-attractive
terms (with $\xi <0$ and $\mathrm{g}<0$) makes the stability segments in the
$N(\mu )$ and $N(\alpha )$ dependences somewhat broader, although generally
similar to those observed in Fig. \ref{fig3} (not shown here in detail).

\subsection{The quadratic nonlinear lattice}

Next we address the system with the quadratic-only nonlinearity, whose local
strength is subjected to periodic modulation that represents a nonlinear
lattice (while the linear potential is absent). This system is modeled by
Eq. (\ref{NLFSE})\ with
\begin{equation}
\xi (x)=-\mathrm{cos}^{2}x,\mathrm{g}=V=0.  \label{nonlin-latt}
\end{equation}%
where the scaling invariance is used to fix the period of the nonlinear
lattice, $L=\pi $, and the center of the soliton may be naturally placed at
a local energy minimum, $x=0$. This setting can be implemented in the LHY
liquid by means of the spatially-periodic Feshbach resonance, taking into
regard the fact that, in unscaled units, $\xi $ in Eq. (\ref{NLFSE}) is
proportional to $\left\vert a\right\vert ^{3/2}$, where $a$ is the
scattering length of the atomic collisions in the binary BEC \cite{Petr-Astr}
(which, in turn, may be modulated in space by means of a spatially-periodic
magnetic field \cite{REV5}).

\begin{figure*}[tbp]
\begin{center}
\includegraphics[width=2\columnwidth]{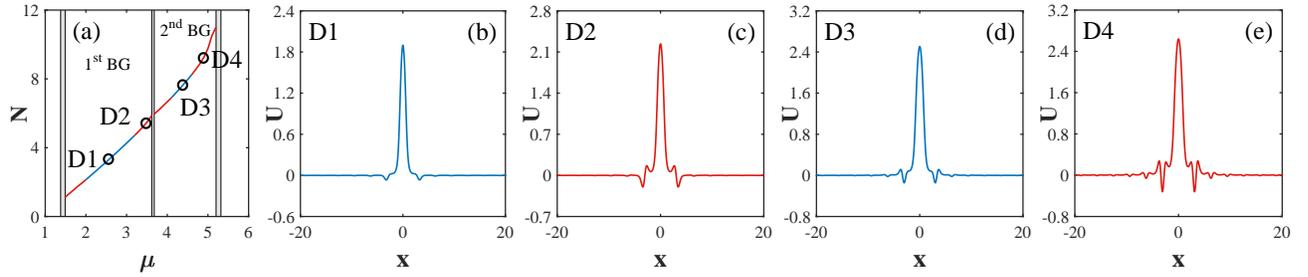}
\end{center}
\caption{(a) The norm $N$ of fundamental GSs supported by the
competing cubic-quintic nonlinearity, versus their chemical potential $%
\protect\mu $. The blue and red segments denote stable and unstable GS
subfamilies, respectively. The profiles of typical GSs: for $\protect\mu =2.6$
(b) $\protect\mu =3.5$ (c); $\protect\mu =4.4$ (d); and $\protect\mu =4.9$
(e). The perturbed evolution of the GSs labeled by D1--D4 is displayed in Figs.
\protect\ref{fig11}(a1--a4), respectively. For all panels, the parameters in Eq. (%
\protect\ref{NLFSE}) and (\protect\ref{LPE}) are $\protect\alpha =1.3$, $%
\protect\xi =-1$, $\mathrm{g}=1$, and $V_{0}=8$.}
\label{fig8}
\end{figure*}

\begin{figure*}[tbp]
\begin{center}
\includegraphics[width=2\columnwidth]{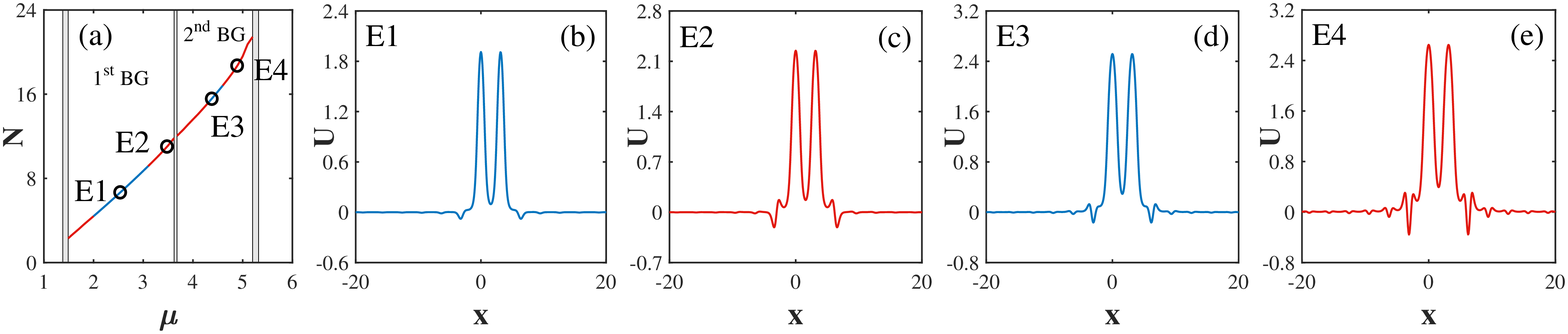}
\end{center}
\caption{The same as in Fig. \protect\ref{fig8}, but for two-peak GSs. Their
profiles are shown for $\protect\mu =2.6$ in (b), $\protect\mu =3.5$ in (c),
$\protect\mu =4.4$ in (d), and $\protect\mu =4.9$ in (e). The perturbed
evolution of the solitons labeled by E1--E4 is displayed in Figs. \protect
\ref{fig11}(b1--b4) respectively. In all panels, the parameters are $\protect%
\alpha =1.3$, $\protect\xi =-1$, $\mathrm{g}=1$, $V_{0}=8$.}
\label{fig9}
\end{figure*}

\begin{figure*}[!tbp]
\begin{center}
\includegraphics[width=2\columnwidth]{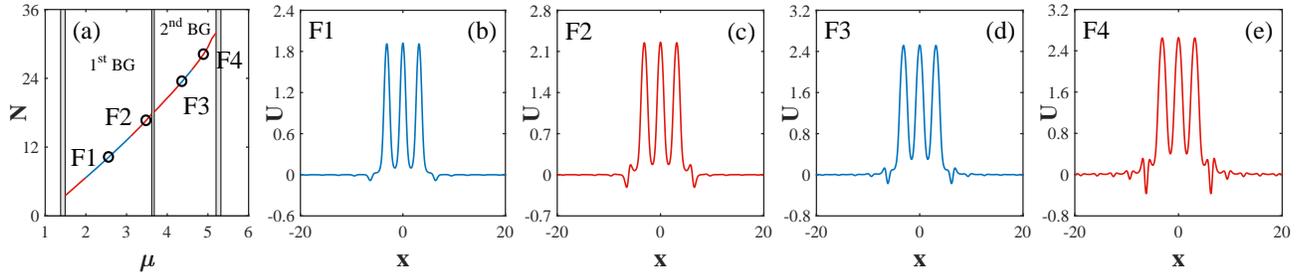}
\end{center}
\caption{The same as in Fig. \protect\ref{fig8}, but for three-peak GSs.
Their profiles are shown for $\protect\mu =2.6$ in (b), $\protect\mu =3.5$
in (c), $\protect\mu =4.4$ in (d), and $\protect\mu =4.9$ in (e). The perturbed
evolution of the solitons labeled by F1--F4 is displayed in Figs. \protect
\ref{fig11}(c1--c4) respectively. In all panels, the parameters are $\protect%
\alpha =1.3$, $\protect\xi =-1$, $\mathrm{g}=1$, $V_{0}=8$.}
\label{fig10}
\end{figure*}

Dependences $N(\mu )$ and $N(\alpha )$ for soliton families produced by the
numerical solution of stationary equation (\ref{NLFSES}), with parameters
taken as per Eq. (\ref{nonlin-latt}), are displayed in Figs. \ref{fig5}%
(a,c). It is worthy to note that, although the spatial modulation of $\xi
(x) $ breaks the scaling relation (\ref{quadr}) between $N$ and $\mu $, the $%
N(\mu )$ curve in Fig. \ref{fig5}(a) almost exactly obeys this relation,
with $\xi (x)$ in Eq. (\ref{nonlin-latt}) replaced by its mean value, $%
\left\langle \xi (x)\right\rangle =1/2$. Furthermore, $N(\mu )$ satisfies
the VK criterion, even if it is far from being a sufficient stability
condition in the present case, since only a relatively small segment of the $%
N(\mu )$ curve is stable in Fig. \ref{fig5}(a).

Particular examples of fundamental (single-peak) solitons supported by the
nonlinear lattice, which are marked by C1--C4 in Figs. \ref{fig5}(a,c), are
presented in Figs. \ref{fig5}(b,d), and their perturbed evolution is
displayed in Figs. \ref{fig6}(a--d). The shape and dynamical behavior of
the solitons are similar to what is shown for their counterparts in the
system with $\xi =-1$ in Figs. \ref{fig1}(b,d) and \ref{fig2}. In
particular, solution C4, shown in Fig. \ref{fig5}(d), which pertains to $%
\alpha =0.5$, represents quasi-Townes solitons in the present case.
Accordingly, it suffers fast decay in Fig. \ref{fig6}(d), similar to the
instability of its counterpart, shown above in Fig. \ref{fig2}(d).

The nonlinear lattice may support, in addition to the single-peak
(fundamental) solitons, also higher-order multi-peak ones. An example of a
two-peak soliton is presented in Fig. \ref{fig5}(e), with separation $\pi $
between the peaks. Both the computation of stability eigenvalues and direct numerical
simulations demonstrate that all higher-order solitons supported by the
nonlinear lattice are subject to instability, accounted for by a pair of
real eigenvalues. In particular, the double-peak soliton shown in Fig. \ref%
{fig5}(e) is associated with unstable eigenvalues $\lambda \approx \pm 0.3$%
. Simulations of the perturbed evolution of this soliton, displayed in Fig. %
\ref{fig6}(e), demonstrate a quick onset of spontaneous symmetry breaking
between the two peaks, which is followed by the development of intrinsic
vibrations of the stronger peak. This dynamical scenario implies that the
attraction between the peaks in the perturbed state is stronger than the
effective pinning induced by the nonlinearity modulation in Eq. (\ref%
{nonlin-latt}).

\section{Gap solitons (GSs) supported by the competing quadratic-cubic (QC)
nonlinearity}
\label{sec4}

\begin{figure*}[tbp]
\begin{center}
\includegraphics[width=1.6\columnwidth]{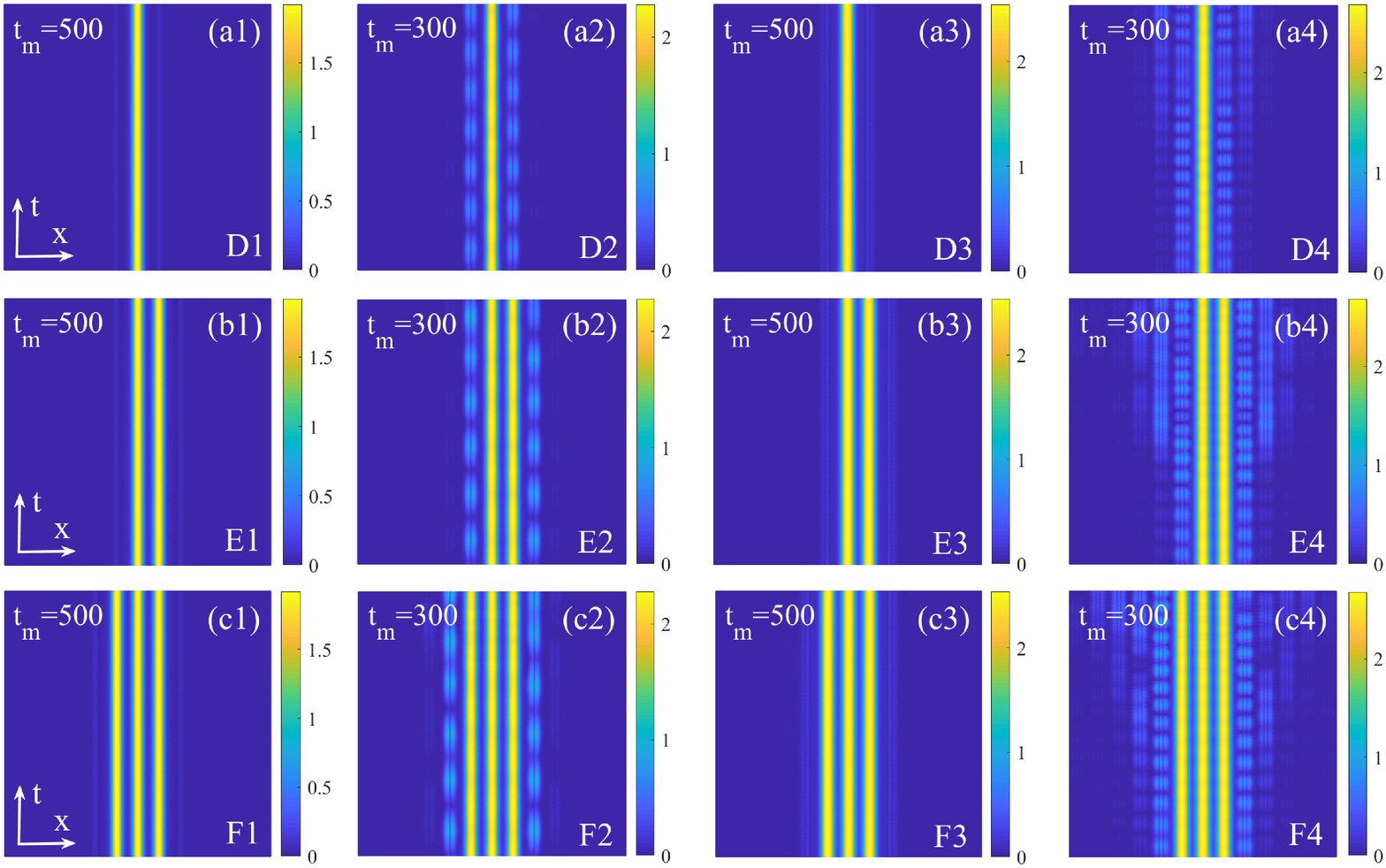}
\end{center}
\caption{The perturbed evolution of fundamental, two-, and tree-peak GSs.
The panels (a1-a4) correspond to the fundamental solitons labeled D1--D4 in Fig.
\protect\ref{fig8}, with values of the chemical potential $\protect\mu =2.6$
(a1), $\protect\mu =3.5$ (a2), $\protect\mu =4.4$ (a3), and $\protect\mu %
=4.9 $ (a4). The panels (b1--b4) correspond to the two-peak solitons labeled
(E1)--(E4) in Fig. \protect\ref{fig9}, with $\protect\mu =2.6$ (b1), $%
\protect\mu =3.5$ (b2), $\protect\mu =4.4$ (b3), and $\protect\mu =4.9$
(b4). The panels (c1--c4) correspond to the three-peak solitons labeled (F1)--(F4)
in Fig. \protect\ref{fig10}, witth $\protect\mu =2.6$ (c1), $\protect\mu =3.5$
(c2), $\protect\mu =4.4$ (c3), and $\protect\mu =4.9$ (c4). The parameters
are $\protect\alpha =1.3$, $\protect\xi =-1$, $\mathrm{g}=1$, and $V_{0}=8$.
In all panels, the evolution is plotted in the interval $-20<x<+20$, while
the values $\mathrm{t}_{\mathrm{m}}$ indicate time intervals in the respective
panels, $0<t<\mathrm{t}_{\mathrm{m}}$.}
\label{fig11}
\end{figure*}

To address families of GSs supported by the lattice potential (\ref{LPE}),
it is first necessary to compute the respective bandgap structure, in the
framework of the linearization of Eq. (\ref{NLFSES}) (a similar computation
was performed in Ref. \cite{Frac5a}). The result is displayed in Fig. \ref%
{fig7}. Note that higher-order bandgaps, which may be populated by GSs,
appears gradually when the LI $\alpha $ decreases from $2$ to $1$, or the
strength $V_{0}$ of the lattice potential increases from $0$ to $15$, as
shown in Figs. \ref{fig7}(a) and (b) respectively. Typical Bloch spectra, in
the form of the chemical potential as a function of the quasi-wavenumber,
obtained at small and large values of $V_{0}$, are displayed in Figs. \ref%
{fig7}(c) and (d), respectively. The former plot features a narrow first
bandgap and has no second one, while in the latter case both the broad first
and second bandgaps are observed in Fig. \ref{fig7}(b).

As mentioned above, GSs produced by the interplay of the self-repulsive or
attractive cubic nonlinearity with a lattice potential in the fractal
setting were considered in Ref. \cite{Frac5a}. It was found that the
repulsive sign of the cubic term gives rise to a family of spatially even
(symmetric) GSs, which are chiefly stable in the first finite bandgap. GSs
of the same type, created by the self-attractive nonlinearity, were
completely unstable, but partly stable GSs with an odd (antisymmetric)
spatial shape might be supported by the attractive cubic term. Here, we
focus on the symmetric GSs, as the most natural species (in particular, it
is straightforward to create it in the experiment, by means of a usual laser
beam with the Gaussian profile) existing under the combined action of the
competing self-attractive quadratic and repulsive cubic terms. Note that the
setting adopted in Ref. \cite{Frac5a} did not allow one to consider
competing nonlinearities.

The family of fundamental GSs is displayed in Fig. \ref{fig8}(a) by means of
its $N(\mu )$ curve, which covers the first two finite bandgaps, and
exhibits alternation of stable (blue) and unstable (red) segments. Note that
the curve satisfies the \textit{anti-VK criterion}, $dN/d\mu >0$, which a
necessary stability condition (but, generally speaking, not a sufficient
one)\ for solitons supported by a self-repulsive nonlinearity (in
particular, for the usual GSs) \cite{AVK}. The alternation of actually
stable and unstable segments is a nontrivial feature of the present GS
family. Qualitatively, it may be construed as a result of the interplay of
the trends to stabilize the GSs by the self-repulsive cubic term, and
destabilize them by the attractive quadratic one.

The profiles of typical fundamental (single-peaked) GSs in the two lowest
bandgaps, labeled by D1--D4 in Fig. \ref{fig8}(a), are plotted in Figs. \ref%
{fig8}(b--e), respectively, showing that the amplitude of such solitons
increases with the growth of $\mu $, and undulations in their shape
(represented by additional side extrema) are pronounced stronger near edges
of the bandgaps than near their centers. Actually, this feature implies that
stable GSs, which are also found closer to the middle of the bandgaps, are
smoother than their unstable counterparts.

Numerically simulated perturbed evolution of these fundamental solitons is
displayed in Figs. \ref{fig11}(a1--a4), respectively. These numerical simulations
suggest that unstable GSs in the first bandgap spontaneously transform into
robust breathers, while in the second bandgap the instability initiates
gradual decay of the solitons.

The underlying lattice potential makes it possible to construct multi-peak
GSs, which may be considered as bound states of single-peak ones. The
resulting family of double-peak GSs is represented in Fig. \ref{fig9}(a) by
the respective $N(\mu )$ dependence covering the two lowest bandgaps and
showing the alternation of stable (blue) and unstable (red) segments. The
comparison of Figs. \ref{fig9}(a) and \ref{fig8}(a) demonstrates that, quite
naturally, the power of the two-peak solitons is almost exactly equal to twice
the power of their single-peak counterparts.

The profiles of typical two-peak solitons, labeled by E1--E4 in Fig. \ref{fig9}%
(a), are shown in Figs. \ref{fig9}(b--e), respectively, and their perturbed
evolution is displayed in Figs. \ref{fig11}(b1--b4). Similar to their
single-peak counterparts shown in Fig. \ref{fig8}, the two-peak GSs develop
stronger undulations near the edge of the bandgaps, in comparison to
smoother shapes near the bandgaps' centers.

The $N(\mu )$ curve for the family of three-peak GSs, also covering the two
lowest bandgaps and separated into stable (blue) and unstable (red)
segments, is presented in Fig. \ref{fig10}(a), where it also satisfies the
anti-VK criterion. Comparing Figs. \ref{fig8}(a), \ref{fig9}(a) and \ref%
{fig10}(a), we conclude that the stability intervals are narrower for both the
two-peak and three-peak GSs, in comparison to those of the fundamental ones.

Typical profiles of the three-peak solitons, marked by F1--F4 in Fig. \ref%
{fig10}(a), are plotted in Figs. \ref{fig10}(b--e) respectively, and their
perturbed evolution is displayed in Figs. \ref{fig11}(c1--c4). Similar to
what is mentioned above for the two-peak GSs, the power of the three-peak ones is
larger almost exactly by a factor of $3$ in comparison to the single-peak
solitons, and the shape undulations of the three-peak GSs are stronger near the
edge of the bandgap. Also similar to the single-peak solitons, those two- and three-peak
modes that are unstable, spontaneously transform into oscillating
states in the first bandgap, and suffer gradual decay in the second one.

\section{Conclusion}
\label{sec5}

The objective of this work was to extend the recently developed models of
nonlinear fractal waveguides for the case of the quadratic and
quadratic-cubic nonlinearities. These models apply to the gas of quantum
particles moving by L\'{e}vy flights, with the self-attractive nonlinearity
representing the LHY correction to the nearly compensated cubic mean-field
nonlinearity in the binary BEC. First, the soliton family was constructed in
the basic case of the purely quadratic nonlinearity (known as the
\textquotedblleft LHY liquid"), and the stability of the solitons was
investigated. In addition to the numerical findings, some essential results
are obtained in the analytical form, such as the $N(\mu )$ dependence, and
the value of the degenerate norm of the \textit{quasi-Townes} soliton
family, at the value of the L\'{e}vy index $\alpha =1/2$. Solitons supported
by the nonlinear lattice, based on the periodically modulated strength of
the quadratic term, were briefly considered too, with a conclusion that only the
fundamental solitons are stable in this case, all multi-peak ones being
unstable.

Next, families of fundamental, double, and triple gap solitons (GSs) were
constructed in the first two finite bandgaps, and their stability
identified, under the action of the competing quadratic and cubic
(attractive and repulsive) terms and lattice potential. It was found that,
due to the presence of the competing nonlinearity, the GS\ branches split
into alternating stable and unstable segments, the solitons tending to be
more stable, with a smoother shape, close to the middle of the bandgaps. The
stability segments are broader for the fundamental (single-peak) GSs than
for the two- and three-peak ones.

This work may be extended in other directions, In particular, it is
interesting to construct moving solitons in the free space (note that the
fractional diffraction breaks the Galilean invariance of the underlying
NLSE, thus making the problem nontrivial), and simulate collisions between
them.

\section*{Funding}

National Major Instruments and Equipment Development Project of National Natural Science Foundation of China (61827815); National Natural Science Foundation of China (62075138); Natural Science Foundation of Guangdong Province (2021A1515011909); Science and Technology Project of Shenzhen (JCYJ20190808121817100, JCYJ20190808164007485, JSGG20191231144201722); Natural Science Foundation of Shenzhen University (2019007); Doctoral Scientific Research Foundation of Jiujiang University (8722509); Israel Science Foundation (grant No. 1286/17).

\section*{Declaration of Competing Interest}

The authors declare that they have no known competing financial interests
or personal relationships that could have appeared to influence the work
reported in this paper.

\end{document}